\documentclass[a4paper]{jpconf}
\usepackage{graphicx,cite,enumerate,amsmath,amsthm,amssymb,hyperref}

\def \BEA { \begin{eqnarray}}
\def \EEA {\end{eqnarray}}
\def \BE {\begin{equation}}
\def \EE {\end{equation}}

\newcommand{\WDS}[1] {\Phi_{#1}^{S}}
\newcommand{\WDA}[1] {\Phi_{#1}^{A}}
\newcommand{\WD}[1] {\Phi_{#1}}

\newcommand{\OO}[1] {{O}(r^{-#1})}
\newcommand{\OOp}[1] {{O}(r^{#1})}

\newcommand{\oo}[1] {o(r^{-#1})}
\newcommand{\oop}[1] {o(r^{#1})}

\newcommand{\R}[1] {\rho_{#1}}		

\newcommand{\Om}[1] {\Omega_{#1}}	
\newcommand{\Ps}[1] {\Psi_{#1}}

\def\a{\alpha}
\def\b{\beta}
\def\g{\gamma}
\def\de{\delta}

\def \bl {\mbox{\boldmath{$\ell$}}}

\def \bn {\mbox{\boldmath{$n$}}}

\def \hbm #1 {\mbox{\boldmath{$\hat m^{(#1)}$}}}
\def \bm {\mbox{\boldmath{$m$}}}

\newcommand{\be}{\begin{equation}}
\newcommand{\ee}{\end{equation}}
\newcommand{\beqn}{\begin{eqnarray}}
\newcommand{\eeqn}{\end{eqnarray}}
\newcommand{\pa}{\partial}
\newcommand{\ba}{\begin{array}}
\newcommand{\ea}{\end{array}}

\begin{document}

\title{Asymptotic properties of gravitational and electromagnetic fields in higher dimensions}

\author{Marcello Ortaggio, Alena Pravdov\' a}

\address{Institute of Mathematics, Academy of Sciences of the Czech Republic \\ \v Zitn\' a 25, 115 67 Prague 1, Czech Republic}

\ead{ortaggio@math.cas.cz, pravdova@math.cas.cz}

\date{\today}

\begin{abstract}

We summarize the fall-off of electromagnetic and gravitational fields in $n>5$ dimensional Ricci-flat spacetimes along an asympotically expanding non-singular geodesic null congruence.

\end{abstract}

\section{Introduction}

\label{sec_intro}

Under suitable assumptions, the well-known peeling-off property characterizes the behavior of the gravitational and electromagnetic fields  at null infinity (see, e.g., \cite{NewTod80,penrosebook2} and references therein). It has been observed \cite{GodRea12} that the Weyl tensor peels off differently in $n>4$ dimensions. Here, we summarize our recent results \cite{OrtPra14,Ortaggio14} on the leading-order behavior of gravitational and electromagnetic fields in higher dimensions. Ref.~\cite{OrtPra14} partly recovers the results of \cite{GodRea12} but uses a different method and different assumptions. We restrict to Ricci-flat spacetimes with suitable properties at null infinity  (a cosmological constant can be included \cite{OrtPra14,Ortaggio14}), formulated in terms of a geodesic null vector field $\bl=\pa_r$ ($r$ is an affine parameter) and of the Weyl tensor, using  a ``null'' frame \cite{OrtPraPra13rev} based on two null vectors $\bm_{(0)}=\bl$, $\bm_{(1)}=\bn$ and $n-2$ orthonormal spacelike vectors $\bm_{(i)}$ ($i, j, \dots=2,\ldots,n-1$). First, we assume that the optical matrix $\R{ij} = \ell_{a;b}m_{(i)}^a m_{(j)}^b$ is {\em asymptotically non-singular and expanding} \cite{OrtPra14,Ortaggio14}
(this includes asymptotically flat spacetimes \cite{GodRea12} but also holds more generally -- see \cite{NP} in four dimensions). Furthermore, we assume that the boost-weight (b.w.) $+2$ Weyl components $\Om{ij}\equiv C_{0i0j}=C_{abcd}\ell^a m_{(i)}^b\ell^c m_{(j)}^d$ fall off as
\be
	\Om{ij}=\OO{\nu} \qquad (\nu>2) .
	\label{Omega}
\ee	
Again, this is satisfied in asymptotically flat spacetimes \cite{GodRea12} (e.g., $\Om{ij}=\OO{5}$ in the 4D spacetimes of \cite{NP}). Under the above conditions, one is able to determine how the Maxwell and Weyl tensors fall off as $r\to\infty$, as we summarize in sections~\ref{sec_em} and \ref{sec_grav}. However, as an intermediate step, one also needs the $r$-dependence of the Ricci rotation coefficients and of the derivative operators \cite{OrtPraPra13rev}, which is given in \cite{OrtPra14} (it follows from the Ricci identities \cite{OrtPraPra07}, also using the commutators \cite{Coleyetal04} and the Bianchi identities \cite{Pravdaetal04}). For example, $\R{ij}=\frac{\delta_{ij}}{r}+\ldots$. For brevity,  in this paper, we discuss  only results in $n>5$ dimensions --  the case $n\ge 5$ is studied in \cite{OrtPra14,Ortaggio14}.

\section{Electromagnetic field}

\label{sec_em}

We start from the simpler case of {\em test} Maxwell fields in the background of an $n$-dimensional Ricci-flat spacetime satisfying the assumptions of section~\ref{sec_intro} \cite{Ortaggio14}. The gravitational field (Weyl tensor) can be treated similarly, however, resulting in a larger number of possible cases (section~\ref{sec_grav}).

In the frame of section~\ref{sec_intro}, we assume that for $r\to\infty$ the Maxwell components have a {\em power-like} behavior described by
\be
 F_{0i}=\OOp{\a} , \qquad F_{01}=\OOp{\b} , \qquad F_{ij}=\OOp{\g} , \qquad F_{1i}=\OOp{\de} . \label{leading}
\ee
The empty-space Maxwell equations $F^a_{\ \; b;a}=0=F_{[ab;c]}$ (see \cite{Durkeeetal10,Ortaggio14} for their GHP and NP form) determine the possible values of $\a$, $\b$, $\g$ and $\de$. We assume that if a generic component $f$ behaves as $f=\OO{\zeta}$ then $\pa_rf=\OO{\zeta-1}$ and $\pa_Af=\OO{\zeta}$. As it turns out, $\a$ can be chosen arbitrarily, giving raise to two main cases, $\a\ge-2$ or $\a<-2$. In the latter, one  needs  to choose whether $\g\ge-2$ or $\g<-2$, and then specify more precisely the value of $\a$, as we detail.

\subsection{Case $\a\ge-2$.} \label{subsubsec_a>=-2}

In this case, all components fall off at the same speed, i.e.,
\be
 F_{0i}=\OOp{\a} , \qquad F_{01}=\OOp{\a} , \qquad F_{ij}=\OOp{\a} , \qquad F_{1i}=\OOp{\a} . 
\ee

The electromagnetic field does not peel. This describes, e.g., a uniform magnetic field permeating asymptotically flat black holes \cite{AliFro04} (or black rings \cite{OrtPra06} if $n=5$ is included, cf.~\cite{Ortaggio14}).

\subsection{Case $\a<-2$.}

Generically, we have 
	\beqn
		& & F_{0i}=\OOp{\a} , \label{+1_gamma=-2} \\
		& & F_{01}=\oo{2} , \qquad F_{ij}=\OO{2} , \label{0_gamma=-2} \\
		& & F_{1i}=\OO{2} \label{-1_gamma=-2} .
	\eeqn
The above behavior includes the special case when $\bl$ is an aligned null direction of the Maxwell field, i.e., $F_{0i}=0$ (in the formal limit $\a\to-\infty$). The leading term is of type II. Examples can be obtained as a ``linearized'' Maxwell field limit of certain full Einstein-Maxwell solutions given in \cite{OrtPodZof08} for even $n$. Several subcases are possible when  $\g<-2$.

\subsubsection{Subcase (a): $\g<-2$ with $1-\frac{n}{2}\le\a<-2$.}

In this case, one has the same results as in section~\ref{subsubsec_a>=-2} above. This subcase does not exist for $n=6$.

\subsubsection{Subcase (b): $\g<-2$ with $-\frac{n}{2}\le\a<1-\frac{n}{2}$.}

Here, we have 
				\beqn
					& & F_{0i}=\OOp{\a} , \label{+1_rad} \\
					& & F_{01}=\OOp{\a} , \qquad  F_{ij}=\OOp{\a} , \label{0_rad} \\
					& & F_{1i}=\OOp{1-n/2} . \label{-1_rad} 
			  \eeqn
The leading term  falls off as $1/r^{\frac{n}{2}-1}$ and is of type N. This is characteristic of radiative fields (note that $T_{11}\propto F_{1i}F_{1i}\sim 1/r^{n-2}$ and the energy flux along $\bl$ can be directly related to the energy loss, at least in the case of asymptotically flat spacetimes -- cf.~\cite{Penrose63,ExtNewPen69,vanderBurg69} for $n=4$). As opposed to the well-known four-dimensional case, here, $\bl$ cannot be aligned with $F_{ab}$ if radiation is present (since $\a\ge-\frac{n}{2}$). In the case $\a=-\frac{n}{2}$, if one assumes that $F_{1i}$ has a power-like behavior also at the subleading order, from the Maxwell equations, one finds 
$F_{1i}=F_{1i}^{(0)}{r^{1-\frac{n}{2}}}+\OO{n/2}$, which gives the peeling-off behavior
		\be
		 F_{ab}=\frac{N_{ab}}{r^{\frac{n}{2}-1}}+\frac{G_{ab}}{r^{\frac{n}{2}}}+\ldots \qquad\left(\a=-\frac{n}{2}\right) .
		\label{peeling_gen}
		\ee 
	The subleading term is algebraically general, which is qualitatively different from the 4D case \cite{NewTod80,penrosebook2,ExtNewPen69,vanderBurg69}. This resembles the behavior of the Weyl tensor of higher dimensional asymptotically flat spacetimes \cite{GodRea12}. See \cite{Ortaggio14} for a possible different peeling-off in five dimensions.

\subsubsection{Subcase (c): $\g<-2$ with $2-n\le\a<-\frac{n}{2}$.}

The same results as in section~\ref{subsubsec_a>=-2} apply.

\subsubsection{Subcase (d): $\g<-2$ with $\a<2-n$.}

We have 
				\beqn
					& & F_{0i}=\OOp{\a} , \label{+1_n>5_Coulomb} \\
					& & F_{01}=\OOp{2-n} , \qquad F_{ij}=\oop{2-n} , \\
					& & F_{1i}=\OOp{2-n} . \label{-1_n>5_Coulomb}
			  \eeqn
The leading term is of type II and falls off as $1/r^{n-2}$ (it is purely electric in the subcase $ F_{1i}=\oop{2-n}$). This behavior includes the Coulomb field of a weakly charged asymptotically flat black hole \cite{AliFro04,Aliev06prd} (or black ring \cite{OrtPra06} if $n=5$ is included \cite{Ortaggio14}). In the special subcase $F_{01}=\oop{2-n}$, the same results as in section~\ref{subsubsec_a>=-2} again apply (for example, for  $n=5$ and $\a=-4$, this is the case of the weak-field limit of the 5D dipole black rings of \cite{Emparan04}).

Let us observe that in all cases, type N fields for which $\bl$ is aligned are not permitted \cite{Ortaggio07,Durkeeetal10}.

\subsection{The case of $p$-forms}

The above results for a 2-form $F_{ab}$ can be extended easily \cite{Ortaggio14} to $p$-form fields satisfying the generalized Maxwell equations (given in \cite{Durkeeetal10} in the GHP notation). In {\em even} dimensions, the special case $p=n/2$ (including  $n=4$, $p=2$) has unique properties. It peels off as
\be
	F_{a_1\ldots a_{p}}=\frac{N_{a_1\ldots a_{p}}}{r^{\frac{n}{2}-1}}+\frac{II_{a_1\ldots a_{p}}}{r^{\frac{n}{2}}}+\ldots \qquad\qquad \left(p=\frac{n}{2}\right) .
	\label{peeling_n/2}
\ee
The (radiative) leading term is of type N and falls off as $1/r^{\frac{n}{2}-1}$. In contrast to the case $p=2$ discussed above (or, in fact, any other $p\neq n/2$), Maxwell fields of type~N aligned with $\bl$ are now permitted \cite{Ortaggio14} and the peeling \eqref{peeling_n/2} applies also in the presence of a cosmological constant \cite{Ortaggio14}. Corresponding solutions of the {\em full Einstein-Maxwell equations} have recently been obtained \cite{OrtPodZof14}.

\section{Gravitational field}

\label{sec_grav}

The method to be used for the Weyl tensor \cite{OrtPra14} is  essentially similar, now $-\nu$ playing the role that $\a$ played above. Instead of the Maxwell equations, one has to integrate the system ``Bianchi-Ricci-commutators''. However, there is now extra freedom in the choice of possible boundary conditions. In particular, three possible choices for the behavior of b.w. +1 components $\Ps{ijk}$ are possible (cases (i), (ii) and (iii) below). Once the fall-off of $\Om{ij}$ and $\Ps{ijk}$ has been specified, the next step is to determine the fall-off of the b.w. 0 components $\WD{ijkl}$
\be
 \WD{ijkl}=\OOp{\beta_c} .
\ee
The parameter $\b_c$ can then be used to label various possible subcases, which we now present.

\subsection{Case (i): $\Om{ij}=\OO{\nu}$, $\Ps{ijk}=\OO{\nu}$}

In all cases given here, we have (this will not be repeated every time below)
\be
	 \Om{ij}=\OO{\nu}  \quad (\nu>2) , \qquad  \Ps{ijk}=\OO{\nu} .
\ee

\subsubsection{Subcase (A): $\b_c=-2$.}

In this case, necessarily $\b_c>-\nu$ and we have the following possible behaviors, depending on how $\nu$ is chosen ({cf.~\cite{OrtPra14} for a few further special subcases)}:
	
	\begin{enumerate}[{A}1:] 
		
	\item 
		\beqn
			& & \WD{ijkl}=\OO{2} , \qquad \WDS{ij}=\oo{2}, \qquad \WDA{ij}=\oo{2}  \qquad (2<\nu\le3) , \nonumber  \\
			& & \Ps{ijk}'=\OO{2} , \\
			& & \Om{ij}'= \OOp{\sigma} \qquad {(-2\le\sigma<-1)} ; \nonumber 
		\eeqn

	\item 
		\beqn
			& & \WD{ijkl}=\OO{2} , \quad \WDS{ij}=\OO{3} , \quad \WD{}=\OO{\nu}, \quad \WDA{ij}=\OO{3}  \quad (\, 3<\nu<4) , \nonumber  \\
			& & \Ps{ijk}'=\OO{2} , \quad \Ps{i}'=\OO{3} , \label{R=0_i_3<nu<4} \\
			& & \Om{ij}'=\OO{2} ; \nonumber 
		\eeqn

	\item 
		\beqn
			& & \WD{ijkl}=\OO{2} , \qquad \WDS{ij}=\OO{3} , \qquad \WD{}=\OO{4} , \qquad \WDA{ij}=\OO{3}  \qquad (\nu{\ge 4}) , \nonumber  \\
			& & \Ps{ijk}'=\OO{2}, \quad \Ps{i}'=\OO{3} , \label{R=0_i_nu>=4} \\
			& & \Om{ij}'=\OO{2} , \nonumber 
		\eeqn
		{with the further restrictions $\WDS{ij}=\OOp{1-\nu}$ for $4\le\nu<5$ and $\WDS{ij}=\OO{4}$ for $\nu\ge5$;}

	\item 		
		\beqn
			& & \WD{ijkl}=\OO{2} , \quad \WDS{ij}=\OOp{1-\nu}, \quad \WD{}=\OO{\nu} , \quad \WDA{ij}=\OO{\nu}  \quad (\nu\ge4, \, \nu\neq n) , \nonumber  \\
			& & \Ps{ijk}'=\OO{2}, \quad \Ps{i}'=\OOp{1-\nu} , \\
			& & \Om{ij}'=\OO{2} ; \nonumber 
		\eeqn

 \item 		
		\beqn
			& & \WD{ijkl}=\OO{2} , \qquad \WDS{ij}=\OOp{1-n}, \qquad \WDA{ij}=\OO{n}  \qquad ({\nu\ge n}) , \nonumber  \\
			& & \Ps{ijk}'=\OO{2}, \quad \Ps{i}'=\OOp{1-n} , \\
			& & \Om{ij}'=\OO{2} . \nonumber 
		\eeqn
				
	None of the above five cases can describe asymptotically flat spacetimes, cf.~\cite{GodRea12}.  In {cases A2--A5}, the leading term  falls off as $1/r^2$ at infinity and it is of type II(abd). In {cases A3--A5}, $\bl$ can be a {multiple} WAND. Examples in case A5 are Robinson-Trautman spacetime \cite{PodOrt06}.

	\end{enumerate}

When $\b_c<-2$, its precise value depends on the value of $\nu$ so that we have to consider the following possible cases.

\subsubsection{Subcase (B): $\b_c<-2$ with $\frac{n}{2}<\nu\le1+\frac{n}{2}$.}

	In this case, $\b_c=-\frac{n}{2}$ and we have
		\beqn
			& & \WD{ijkl}=\OO{n/2} , \qquad \WD{}=\OO{\nu} , \qquad \WDA{ij}=\OO{\nu}  \qquad \left(\frac{n}{2}<\nu\le1+\frac{n}{2}\right) , \nonumber  \\
			& & \Ps{ijk}'=\OO{n/2} , \label{R=0_i_beta=-n/2} \\
			& & \Om{ij}'=\OOp{1-n/2} . \nonumber 
		\eeqn	
 Here, $\bl$ cannot be a WAND. The leading term at infinity falls off as $1/r^{n/2-1}$ and it is of type N. This includes {\em radiative spacetimes} \cite{GodRea12} that are asymptotically flat in the Bondi definition \cite{TanTanShi10,TanKinShi11}. If one takes for b.w. +2 components $\nu=1+\frac{n}{2}$ and additionally {\em assumes} that $\Om{ij}=\Om{ij}^{(0)}r^{-n/2-1}+\Om{ij}^{(1)}{r^{-n/2-2}}+\oop{-n/2-2}$, then one finds \cite{OrtPra14} the peeling-off behavior
				\be
					C_{abcd}=\frac{N_{abcd}}{r^{n/2-1}}+\frac{II_{abcd}}{r^{n/2}}+\oo{n/2}  .
					\label{peel_R=0_i}
				\ee
This agrees with \cite{GodRea12} for asymptotically flat spacetimes. See \cite{GodRea12,OrtPra14} for special properties of the case $n=5$. When $\b_c<-2$ but $\nu$ is not in the range $\frac{n}{2}<\nu\le1+\frac{n}{2}$ one has the following subcases (B*) and (C).

\subsubsection{Subcase (B*): $\b_c<-2$  with $2<\nu\le\frac{n}{2}$ or $1+\frac{n}{2}<\nu\le n-1$.}	In this case, $\b_c=-\nu$ and we have	(cf. section~IV~A~5 of \cite{OrtPra14})		
\beqn
	& & \WD{ijkl}=\OO{\nu}, \qquad \WDA{ij}=\OO{\nu},   \nonumber \\
	& & \Ps{ijk}'=\OO{2} \quad\mbox{if } 2<\nu\le3, \qquad \Ps{ijk}'=\OO{\nu} \quad\mbox{if } \nu>3, \label{R=0_generic} \\
	& & \Om{ij}'=\oop{1-\nu}  \quad\mbox{if } \nu\neq\frac{n}{2}, \qquad \Om{ij}'=\OOp{1-n/2}  \quad\mbox{if } \nu=\frac{n}{2}. \nonumber 
\eeqn
Here, $\bl$ cannot be a WAND.

\subsubsection{Subcase (C): $\b_c<-2$ with $\nu>n-1$.}	In this case, $\b_c=1-n$ and we have

	\beqn
			& & \WD{ijkl}=\OOp{1-n} , \qquad \WDA{ij}=\oop{1-n}  \qquad (\nu>n-1) , \nonumber  \\
			& & \Ps{ijk}'=\OOp{1-n} , \label{R=0_i_betac_=1-n} \\
			& & \Om{ij}'=\oop{2-n} , \nonumber 
		\eeqn	 
	with $\WDA{ij}=\OO{\nu}$ for $n-1<\nu<n$ and $\WDA{ij}=\OO{n}$ for $\nu\ge n$. Here, $\bl$ can {become a multiple WAND, cf.~\cite{OrtPraPra09b}.} This includes asymptotically flat spacetimes in the case of {\em vanishing radiation} \cite{GodRea12}, such as  those for which $\bl$ is a multiple WAND \cite{OrtPraPra09b}, e.g., the Schwarzschild-Tangherlini metric and Kerr-Schild spacetimes \cite{OrtPraPra09} with a non-degenerate Kerr-Schild vector.

\subsection{Case (ii): $\Om{ij}=\oo{n}$, $\Ps{ijk}=\OO{n}$}

\subsubsection{Subcase $\b_c=-2$.} Generically, one has
\beqn
	& & \Om{ij}=\oo{n} , \nonumber \\	
	& & \Ps{ijk}=\OO{n} , \nonumber \\
	& & \WD{ijkl}=\OO{2}, \qquad \WDS{ij}=\OO{4} , \qquad \WDA{ij}=\OO{3} ,  \label{R=0_ii_betac=-2} \\
	& & \Ps{ijk}'=\OO{2} , \qquad \Ps{i}'=\OO{3} , \nonumber \\
	& & {\Om{ij}'=\OO{2}} . \nonumber 
\eeqn
For $\Ps{ijk}^{(n)}=0$, this case reduces to \eqref{R=0_i_nu>=4} (with $\nu>n$). See \cite{OrtPra14} for possible subcases.

\subsubsection{Subcase $\b_c=1-n$.} When $\b_c<-2$ then necessarily $\b_c=1-n$ and generically, one has
\beqn
	& & \Om{ij}=\oo{n} , \nonumber \\
	& & \Ps{ijk}=\OO{n} , \nonumber \\
  & & \WD{ijkl}=\OOp{1-n} , \qquad \WDA{ij}=\OO{n} ,  \label{R=0_ii_beta_c=1-n} \ \\	
	& & \Ps{ijk}'=\OOp{1-n} , \qquad \Ps{i}'=\OOp{1-n}   ,\nonumber \\
	& & \Om{ij}'=\oop{2-n} . \nonumber 
\eeqn
This includes asymptotically flat spacetimes in the case of vanishing radiation \cite{GodRea12}. For $\Ps{ijk}^{(n)}=0$, this case reduces to \eqref{R=0_i_betac_=1-n} (with $\nu>n$).

\subsection{Case (iii): $\Om{ij}=\oo{3}$, $\Ps{ijk}=\OO{3}$}

This case cannot represent asymptotically flat spacetimes \cite{GodRea12}. Generically, $\b_c=-2$ and
\beqn
	& & \Om{ij}=\OO{\nu} \qquad (\nu>3) , \nonumber \\
	& & \Ps{ijk}=\OO{3} , \qquad \Ps{i}=\oo{3} , \nonumber \\
	& & \WD{ijkl}=\OO{2}, \qquad \WDS{ij}=\OO{3} , \qquad \WD{}=\oo{3} , \qquad \WDA{ij}=\OO{3} ,  \\
	& & \Ps{ijk}'=\OO{2} , \qquad \Ps{i}'=\OO{3}  ,\nonumber \\
	& & {\Om{ij}'=\OO{2}} , \nonumber 
\eeqn
where $\Ps{i}=\OO{\nu}$, $\WD{}=\OO{\nu}$ for $3<\nu\le 4$ while $\Ps{i}=\OO{4}$, $\WD{}=\OO{4}$ for $\nu>4$. Here, $\bl$ can be a single WAND and the asymptotically leading term is of type II(abd). For $\Ps{ijk}^{(3)}=0$, this case reduces for $3<\nu<4$ to \eqref{R=0_i_3<nu<4} (with $\nu>n$), for $4\le\nu\le n$ to \eqref{R=0_i_nu>=4} and for $\nu>n$ to \eqref{R=0_ii_betac=-2}. If $\b_c<-2$ then $\WD{ijkl}=\OO{3}$ and the leading term at infinity becomes of type III(a).

\ack 

The authors acknowledge support from research plan {RVO: 67985840} and research grant GA\v CR 13-10042S.

\section*{References}

%
%

\providecommand{\newblock}{}

\end{document}